\documentclass[]{article}

\usepackage{cite}
\usepackage{amsmath,amssymb,amsfonts,nccmath,nicefrac}
\usepackage{mathtools}
\usepackage{algorithmic}
\usepackage{graphicx}
\usepackage{textcomp}
\usepackage{adjustbox}
\usepackage{subcaption}
\usepackage{comment}
\usepackage{amssymb}
\usepackage{lipsum}
\usepackage{balance}
\usepackage[belowskip=0pt]{caption}
\usepackage[margin=1.75in]{geometry}

\title{Rapid TAURUS for Relaxation-Based Color Magnetic Particle Imaging}
\author{M. Tun\c{c} Arslan, A. Alper \"{O}zaslan, Semih Kurt, Yavuz Muslu \\ and Emine Ulku Saritas \thanks{This work was supported by the Scientific and Technological Council of Turkey (TUBITAK) under Grants 120E208 and 217S069.}
	\thanks{M. T. Arslan, A. A. \"{O}zaslan, and E. U. Saritas  are with the Department of
		Electrical and Electronics Engineering, Bilkent University, 06800 Ankara,
		Turkey, and also with the National Magnetic Resonance Research
		Center (UMRAM), Bilkent University, 06800 Ankara, Turkey. E. U. Saritas is also with the Neuroscience Program,
		Sabuncu Brain Research Center, Bilkent University, 06800 Ankara,
		Turkey (corresponding author e-mail:
		mtarslan@ee.bilkent.edu.tr).}
	\thanks{Y. Muslu is with the Department of Biomedical Engineering, University of Wisconsin‐Madison, Madison, WI, USA, and also with the Department of Radiology, University of Wisconsin‐Madison, Madison, WI, USA.}
	\thanks{S. Kurt is with the Department of Electrical Engineering and Computer Science, KTH Royal Institute of Technology, Stockholm, Sweden, and also with the Science for Life Laboratory, Stockholm, Sweden.}
}

\date{}

\begin{document}

\maketitle

\begin{abstract}
Magnetic particle imaging (MPI) is a rapidly developing medical imaging modality that exploits the non-linear response of magnetic nanoparticles (MNPs). Color MPI widens the functionality of MPI, empowering it with the capability to distinguish different MNPs and/or MNP environments. The system function approach for color MPI relies on extensive calibrations that capture the differences in the harmonic responses of the MNPs. An alternative calibration-free x-space-based method called TAURUS estimates a map of the relaxation time constant, $\tau$, by recovering the underlying mirror symmetry in the MPI signal. However, TAURUS requires a back and forth scanning of a given region, restricting its usage to slow trajectories with constant or piecewise constant focus fields (FFs). In this work, we propose a novel technique to increase the performance of TAURUS and enable $\tau$ map estimation for rapid and multi-dimensional trajectories. The proposed technique is based on correcting the distortions on mirror symmetry induced by time-varying FFs. We demonstrate via simulations and experiments in our in-house MPI scanner that the proposed method successfully estimates high-fidelity $\tau$ maps for rapid trajectories that provide orders of magnitude reduction in scanning time, while preserving the calibration-free property of TAURUS.
\end{abstract}

\section{Introduction}
\label{sec:introduction}
Magnetic particle imaging (MPI) is a rapidly developing tracer-based medical imaging modality, which takes advantage of the non-linear response of magnetic nanoparticles (MNPs) under time-varying magnetic fields  \cite{gleich2005tomographic,goodwill2012x,saritas2013magnetic,zheng2017seeing,wu2019review}. The leading applications of MPI include angiography \cite{molwitz2019first,mohtashamdolatshahi2020vivo}, stem cell tracking \cite{zheng2015magnetic,them2016increasing,wang2020artificially}, inflammation imaging \cite{mangarova2020ex,chandrasekharan2021non}, and localized hyperthemia \cite{tay2018magnetic}. Color MPI is an emerging field within MPI, providing a means to distinguish different MNPs and/or MNP environments \cite{rahmer2015first}. Color MPI offers various practical applications such as catheter tracking during cardiovascular interventions \cite{haegele2016multi,rahmer2017interactive}, and identifying the characteristics of the environment such as viscosity \cite{utkur2017relaxation,shasha2017harmonic,moddel2018viscosity,utkur2019relaxation,draack2021magnetic} and temperature \cite{stehning2016simultaneous,zhong2018magnetic,shi2019concurrent}. 

The first color MPI study targeted distinguishing MNPs using a system function reconstruction (SFR) approach \cite{rahmer2015first}. SFR requires lengthy calibration measurements of a point source MNP at all voxel locations within a field-of-view (FOV)
\cite{rahmer2009signal,rahmer2012analysis}.
For the color MPI extension, a separate calibration is needed for each MNP type and/or environment parameter, further lengthening the calibration time \cite{rahmer2015first, haegele2016multi,szwargulski2020monitoring,stehning2016simultaneous}.


X-space-based color MPI techniques do not require a calibration, but the resulting images are typically blurred by the point spread function (PSF) of the imaging system \cite{goodwill2010x}. One approach is performing multiple measurements at different drive field (DF) amplitudes to differentiate the relaxation behaviors of MNPs \cite{hensley2015preliminary}. Another approach is a relaxation time constant ($\tau$) estimation method, abbreviated as TAURUS ($\tau$ estimation via Recovery of Underlying mirror Symmetry) \cite{utkur2017relaxation,utkur2019relaxation,muslu2018calibration}. TAURUS estimates a $\tau$ for each partial field-of-view (pFOV) (or patch) within the image to create a quantitative $\tau$ map. Importantly, it does not require any calibrations, multiple measurements, or prior knowledge about the MNPs to estimate the $\tau$ map. However, TAURUS relies on the underlying mirror symmetry of the MPI signal, which is valid only for trajectories that perform a strict back-and-forth scanning of each pFOV. This requirement restricts the usage of TAURUS to trajectories with constant or piecewise constant focus fields (FFs), with long scanning times. 

In this work, we propose a novel technique that enables $\tau$ map estimation via TAURUS for rapid and multi-dimensional trajectories that utilize time-varying FFs. We first demonstrate how the time-varying FFs distort the mirror symmetry of the MPI signal, and propose a method to compensate for the FF-induced distortions. Additionally, we improve the performance of TAURUS by casting it as a weighted least squares (WLS-TAURUS) problem. With simulations and imaging experiments, we demonstrate that the proposed method is robust for a wide range of FF slew rates (SRs) and noise, successfully estimating $\tau$ maps for rapid multi-dimensional trajectories, while preserving the calibration-free property of TAURUS. The results show that the proposed method provides high fidelity $\tau$ maps and orders of magnitude reduction in total scanning time for TAURUS.

\section{Theory}
\subsection{Mirror Symmetry and TAURUS}
In MPI, the relaxation effect is modeled as a first-order Debye process, expressed as a convolution relation  \cite{croft2012relaxation}:
\begin{subequations}
	\label{eq:nonadiabatic_signal} 
	\begin{gather}
		s(t) = s_{\text{adiab}} (t) * r_{\tau}(t),  \\
		r_{\tau}(t) = \frac{1}{\tau} e^{-t/\tau} u(t).
	\end{gather} 
\end{subequations}
Here, $*$ denotes convolution, $s(t)$ is the signal with relaxation, $s_{\text{adiab}} (t)$ is the adiabatic signal described by the Langevin model \cite{goodwill2010x}, $r_{\tau}(t)$ is the relaxation kernel, $\tau$~(s) is the effective relaxation time constant of the MNP that explains the signal lag \cite{croft2012relaxation}, and $u(t)$ is the Heaviside unit step function.

TAURUS uses the underlying mirror symmetry of the adiabatic MPI signal to estimate the time constant $\tau$ \cite{utkur2017relaxation,muslu2018calibration}. The mirror symmetry assumption is valid independent of the MNP type or distribution in space, as long as the signal is acquired during a back-and-forth FFP trajectory. Without loss of generality, consider the following trajectory with a 1D DF in the z-direction together with constant FFs:
\begin{equation}
	\boldsymbol{x}_s(t) = \begin{bmatrix} x(t)\\ y(t) \\ z(t) \end{bmatrix} = \underset{\text{Focus Field}}{\underbrace{\begin{bmatrix} \nicefrac{B_{F,x}}{G_x} \\ \nicefrac{B_{F,y}}{G_y}  \\ \nicefrac{B_{F,z}}{G_z} \end{bmatrix}}} + \underset{\text{Drive Field}}{\underbrace{\begin{bmatrix} 0 \\ 0 \\ \frac{B_p}{G_z}cos(2\pi f_d t) \end{bmatrix}}}
	\label{eq:general_trajectory}.
\end{equation}
Here, $B_{F,i}$~(T) are the constant FFs and $G_i$~(T/m) are the selection field (SF) gradients along each direction, respectively. In addition, $B_p$~(T) and $f_d$~(Hz) are the amplitude and the frequency of the DF, respectively. The size of pFOV covered by the DF alone is equal to $W_p=2 B_p/G_z$ \cite{saritas2013magnetic}. We refer to this trajectory as the ``piecewise trajectory" (PWT), as it utilizes piecewise constant FFs to scan a small pFOV. Then, $B_{F,i}$ are stepped to different values to cover the entire FOV. 

For PWT, we define the negative and positive signals as the half-cycle signals acquired during the back and forth portions of the FFP movement, respectively. Then, the mirror symmetry in $s_{\text{adiab}} (t)$ can be expressed as \cite{utkur2017relaxation,muslu2018calibration}:
\begin{equation}
	s_{\text{pos,adiab}}(t) = -s_{\text{neg,adiab}}(-t) = s_{\text{half}}(t).
	\label{eq:half_signal}
\end{equation}
For the signal with relaxation, however, this mirror symmetry is broken. Using Eqs.~\eqref{eq:nonadiabatic_signal}~and~\eqref{eq:half_signal}, we can write:
\begin{subequations}
	\label{eq:time_domain_eqs}
	\begin{gather}
		s_{\text{pos}}(t) = s_{\text{half}}(t) * r_{\tau}(t) \\
		s_{\text{neg}}(t) = -s_{\text{half}}(-t) * r_{\tau}(t). 
	\end{gather}
\end{subequations}
The Fourier transforms of these signals and $r_{\tau}(t)$ are:
\begin{subequations}
	\label{eq:frequency_domain_eqs}
	\begin{align}
		S_{\text{pos}}(f) & = \mathcal{F} \{s_{\text{pos}} (t)\} = S_{\text{half}} (f) R_{\tau}(f), \\
		S_{\text{neg}}(f) & = \mathcal{F} \{s_{\text{neg}} (t)\} = -S^*_{\text{half}} (f) R_{\tau}(f), \\
		R_{\tau}(f) & = \mathcal{F}\{r_{\tau}(t) \} = \frac{1}{1 + i2\pi f\tau}.
	\end{align}
\end{subequations}
Here, $\mathcal{F}$ denotes Fourier transformation. Next, $\tau$ can be computed in frequency domain as follows \cite{utkur2017relaxation,muslu2018calibration}:
\begin{equation}
	\tau (f) = \frac{S^*_{\text{pos}} (f) + S_{\text{neg}} (f)}{i2\pi f \big(S^*_{\text{pos}} (f) - S_{\text{neg}} (f) \big)}.
	\label{eq:frequency_division_tau}
\end{equation}
Here, superscript $*$ denotes complex conjugation. Ideally, $\tau (f)$ should be independent of frequency, $f$. However, the presence of noise or deviations from the model in  Eq.~\eqref{eq:nonadiabatic_signal} can cause frequency dependency. To provide robustnesss against such non-idealities, a weighted average of $\tau(f)$ is computed using the magnitude spectrum $|S_{\text{pos}} (f)|$ as weights \cite{muslu2018calibration}.
Once $\tau$ is computed, the mirror symmetric $s_{\text{adiab}} (t)$ can be recovered by deconvolving $s(t)$ with $r_{\tau}(t)$. 

\begin{figure}[t]
	\centering
	\includegraphics[width=0.8\linewidth]{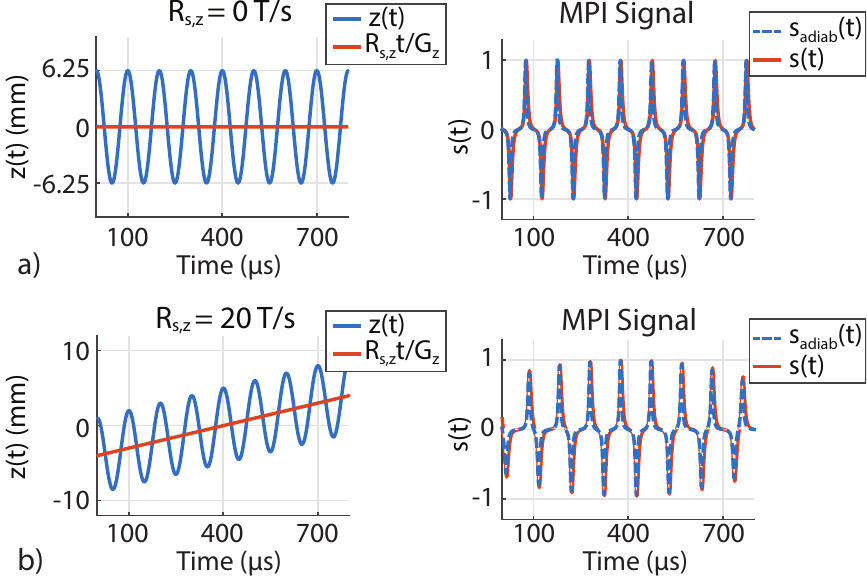}
	\caption{Effects of a linearly ramping FF, demonstrating a time-varying signal amplitude at non-zero $R_{s,z}$. The FFP position, and the adiabatic and non-adiabatic signals for (a) $R_{s,z}=0$ and (b) $R_{s,z}=20$~T/s.  
		Simulations utilized a point source MNP with $\tau = 3~\mu$s positioned at $z = 0$.
	}
	\label{fig:recovery_plot_trajectory_signal}
\end{figure}

\subsection{FF-Induced Distortions in Mirror Symmetry}
Trajectories that contain time-varying FFs experience a distortion in mirror symmetry, even for the case of $s_{\text{adiab}} (t)$. Without loss of generality, consider the following FFP trajectory, with a 1D DF and a linearly ramping FF applied along the z-direction, and constant FFs in the x- and y-directions:
\begin{equation}
	\boldsymbol{x}_s(t) = \begin{bmatrix} x(t)\\ y(t) \\ z(t) \end{bmatrix} = \begin{bmatrix} \nicefrac{B_{F,x}}{G_x} \\ \nicefrac{B_{F,y}}{G_y} \\ \nicefrac{R_{s,z}t}{G_z} \end{bmatrix} + \begin{bmatrix} 0 \\ 0 \\ \frac{B_p}{G_z}cos(2\pi f_d t) \end{bmatrix}
	\label{eq:trajectory}.
\end{equation}
Here, $R_{s,z}$~(T/s) is the SR of the FF along the z-direction. As shown in Fig.~\ref{fig:recovery_plot_trajectory_signal}, 
the positive and negative signals have matching amplitudes for $R_{s,z} = 0$. For $R_{s,z} = 20$~T/s, on the other hand, the amplitude is time varying due to the global FFP movement caused by the FF. The pFOV center moves by $R_{s,z}/(2f_d G_z)$ in half a DF period, so the FFP motion is no longer symmetrical around the pFOV center. In addition, the trajectory speed is different for the positive and negative signals, which further distorts the  mirror symmetry in $s_{\text{adiab}} (t)$.

\begin{figure}[t]
	\centering
	\includegraphics[width=0.8\linewidth]{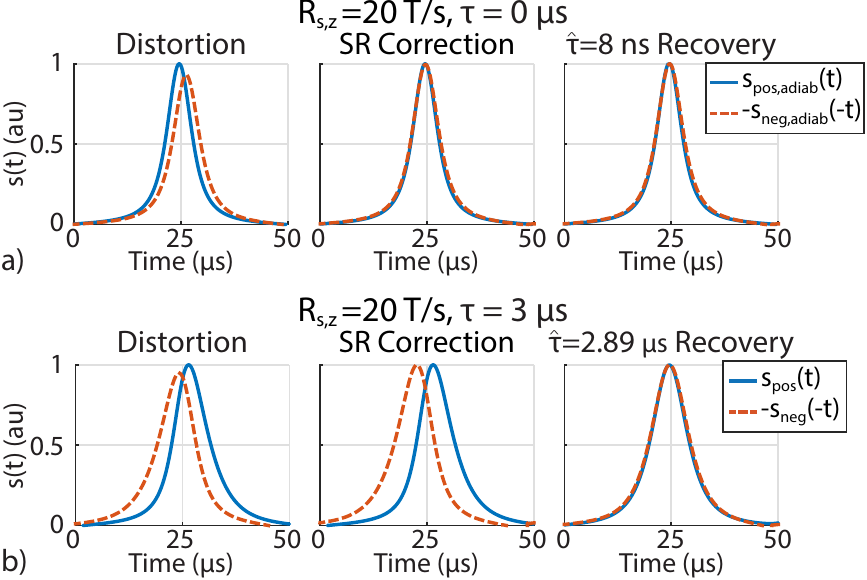}
	\caption{FF-induced distortions on mirror symmetry at $R_{s,z}=20$~T/s for (a) the adiabatic (i.e., $\tau = 0$) and (b) the non-adiabatic (for $\tau = 3~\mu$s) signals. After SR correction, the mirror symmetry is recovered for the adiabatic signal. After SR correction, TAURUS yields (a) $\hat{\tau} = 8$~ns and (b) $\hat{\tau} = 2.89~\mu$s.
	}
	\label{fig:recovery_plot_distortion_correction}
\end{figure}

In Fig.~\ref{fig:recovery_plot_distortion_correction}.(a), the positive and mirrored negative signals for $s_{\text{adiab}} (t)$ (i.e., $\tau = 0$) are plotted, showing the FF-induced distortions on mirror symmetry. The SR along the z-axis causes a lag between these two signals and a mismatch between their amplitudes. 

\subsection{SR Correction for Recovering Mirror Symmetry}
For accurate $\tau$ estimations, we propose an SR correction method to correct the FF-induced distortions and recover the mirror symmetry in $s_{\text{adiab}} (t)$. First, we assume that the FFP speed is dominated by the DF for the majority of the FFP motion. For Eq.~\eqref{eq:trajectory}, this assumption can be expressed as: 
\begin{subequations}
	\label{eq:basic_condition}
	\begin{align}
		\text{max} \Big\{\Big| \frac{d}{dt} \Big( \frac{B_p}{G_z}cos(2\pi f_d t) \Big)\Big|\Big\} & \gg \Big|\frac{d}{dt} \Big(\frac{R_{s,z}}{G_z}t \Big)\Big|, \\
		B_p 2\pi f_d & \gg R_{s,z}.   
	\end{align}
\end{subequations}
Under this assumption, the overall effect of the FF can be approximated as a global time shift and a global amplitude scaling between the positive and negative signals. 

A closed-form expression for the FF-induced time shift, $\Delta t$, can be found by solving the following equation:
\begin{subequations}
	\begin{gather}
		z(t_0) = z(t_0 + T/2 + \Delta t), \label{eq:position_eq} \\ 
		t_0  = \underset{t}{\text{argmax}}(|\Dot{z}(t)|), \enspace t \in [0, T/2].
	\end{gather}
\end{subequations}
Here, $t_0$ is the time point during the first negative half-cycle at which the FFP speed is maximum. 
Without loss of generality, consider the trajectory in Eq.~\eqref{eq:trajectory}, where $t_0$ can be found as $1/(4f_d)$. Inserting this $t_0$ into Eq.~\eqref{eq:position_eq} and simplifying yields:
\begin{equation}
	B_p sin(2\pi f_d \Delta t) + R_{s,z} \Delta t + \frac{R_{s,z}}{2f_d} = 0.
	\label{eq:deltat}
\end{equation}
If $\Delta t \ll 1 / f_d$, a first-order Taylor series expansion of sine yields $\Delta t = -R_{s,z} / (2 f_d) / (B_p 2 \pi f_d + R_{s,z})$. A more accurate solution can be computed, e.g., using a higher-order Taylor series expansion of sine.

Next, the amplitude scaling, $\alpha_{\Delta t}$, can be expressed as the ratio of the FFP speeds at $t_0$ and $t_0 + T/2 + \Delta t$, which yields:
\begin{equation}
	\alpha_{\Delta t} = \frac{|B_p 2\pi f_d cos(2 \pi f_d \Delta t) + R_{s,z}|}{|-B_p 2\pi f_d + R_{s,z}|}.
	\label{eq:alphadeltat}
\end{equation}
Once $\Delta t$ and $\alpha_{\Delta t}$ are computed using Eqs.~\eqref{eq:deltat}-\eqref{eq:alphadeltat}, the positive signal can be kept the same and the proposed SR correction can be applied to the negative signal only, i.e., 
\begin{subequations}
	\label{eq:corrected}
	\begin{align}
		& S_{\text{pos,c}}(f) = S_{\text{pos}}(f), \\
		& S_{\text{neg,c}}(f) = S_{\text{neg}}(f)e^{i2\pi \Delta t f}\alpha_{\Delta t}. 
	\end{align}
\end{subequations}
Finally, Eq.~\eqref{eq:frequency_division_tau} can be modified as follows:
\begin{equation}
	\tau (f) = \frac{S^*_{\text{pos,c}} (f) + S_{\text{neg,c}} (f)}{i2\pi f \big(S^*_{\text{pos,c}} (f) - S_{\text{neg,c}} (f) \big)}.
	\label{eq:frequency_division_tau_corrected}
\end{equation}

TAURUS was originally proposed for the case of PWT, 
for which $R_{s,z} = 0$ and Eqs. \eqref{eq:deltat}-\eqref{eq:alphadeltat} yield $\Delta t = 0$ and $\alpha_{\Delta t} = 1$, as expected. As $R_{s,z}$ increases, both $\Delta t$ and $\alpha_{\Delta t}$ increase steadily. Importantly, both $\Delta t$ and $\alpha_{\Delta t}$ are independent of the MNP parameters or the environmental conditions, and are purely dependent on the system and trajectory parameters.

\begin{figure}[t]
	\centering
	\includegraphics[width=0.8\linewidth]{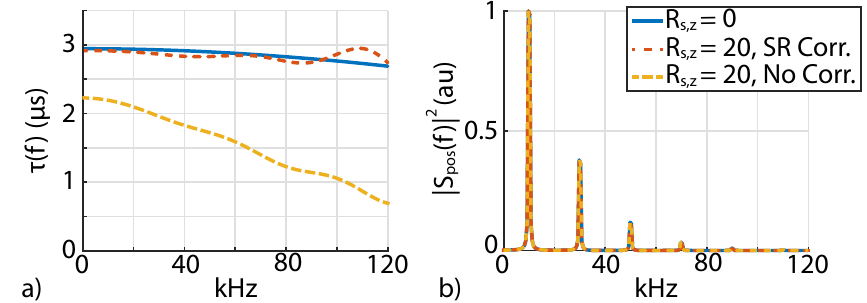}
	\caption{ 
		(a) $\tau (f)$ and (b) normalized power spectrum before and after SR correction for $R_{s,z} = 20$~T/s. The SR-corrected $\tau (f)$ closely follows the reference case of $R_{s,z} = 0$, while the non-corrected $\tau (f)$ shows increasing levels of underestimation at higher frequencies. Power spectra match closely. Simulations utilized a point source MNP with $\tau = 3~\mu$s.
	}
	\label{fig:tau_f_sf_squared_graph}
\end{figure}

In Fig.~\ref{fig:recovery_plot_distortion_correction}, the positive and negative signals of $s_{\text{adiab}}(t)$ (i.e., $\tau = 0$) and $s(t)$ (for $\tau = 3~\mu$s) are shown for $R_{s,z}=20$~T/s. For accurate $\tau$ estimations, $s_{\text{adiab}}(t)$ should have a perfect mirror symmetry. However, due to the FF-induced distortions, the positive and the mirrored negative signals of $s_{\text{adiab}}(t)$ are shifted in opposite directions in time in Fig.~\ref{fig:recovery_plot_distortion_correction}.(a). Additionally, the amplitude of the negative signal is smaller than that of the positive signal. When SR correction is applied using Eq.~\eqref{eq:corrected}, the mirror symmetry is recovered for $s_{\text{adiab}}(t)$. Computing $\tau$ using TAURUS yields $\hat{\tau} = 8$~ns in this case (i.e., practically zero, as expected). 
For $s(t)$ in Fig.~\ref{fig:recovery_plot_distortion_correction}.(b), the relaxation effect causes a separation of the positive and mirrored negative signals in the direction opposite to that caused by the SR. The relaxation effect also broadens the signal along the scanning direction. Directly computing $\tau$ using this distorted signal yields $\hat{\tau} = 1.85~\mu$s, a significant underestimation. After applying SR correction, the amplitudes of the positive and negative signals match and the effective delay between them gets visibly larger. Then, computing $\tau$ for the SR-corrected signal gives $\hat{\tau} = 2.89~\mu$s, a very close match to the actual value. Finally, applying deconvolution using this $\hat{\tau}$ recovers the underlying mirror symmetry of the MPI signal. 

In Fig.~\ref{fig:tau_f_sf_squared_graph}, $\tau (f)$ and power spectrum are shown before and after SR correction for $R_{s,z} = 20$~T/s. Here, the plots for $R_{s,z} = 0$ are provided as the references without any FF-induced distortions. As seen in Fig.~\ref{fig:tau_f_sf_squared_graph}.(a), the SR-corrected $\tau (f)$ closely follows the reference, whereas the non-corrected $\tau (f)$ shows increasing levels of underestimation at higher frequencies. Here, the ripple effect on $\tau (f)$ at $R_{s,z} = 20$~T/s is caused by the digital manipulations and temporal windowing of the signal. 
If the FFP speed is strongly dominated by the DF (see Eq.~\eqref{eq:basic_condition}), 
these ripples flatten and SR-corrected $\tau (f)$ converges to the reference (results not shown). Nevertheless, since TAURUS applies an averaging of $\tau (f)$ weighted by the magnitude spectrum, the ripples in $\tau (f)$ at high frequencies have a relatively minor effect on the final estimated $\tau$. 
In Fig.~\ref{fig:tau_f_sf_squared_graph}.(b), the normalized power spectrum of $s_{pos}(t)$ is shown, where there is negligible difference between the SR-corrected and non-corrected cases and the reference. 

\subsection{Rapid Color MPI Trajectories}
The trajectory in Eq.~\eqref{eq:trajectory} covers the FOV as a ``line-by-line trajectory" (LLT), with overlapping pFOVs along the z-direction and discrete steps in the x- and y-directions. Due to the continuously ramping FF, this trajectory is already much faster than PWT. For example, for the FFP scanner used in this work with $G_z = 2.4$~T/m, even a relatively modest SR of $R_{s,z} = 2$~T/s corresponds to 0.83 m/s speed, which can cover the human height in approximately 2 seconds.

To cover a 2D or 3D FOV continuously, time-varying FFs, $B_{F,i}(t)$, can be utilized in all directions, i.e.,
\begin{equation}
	\boldsymbol{x}_s(t) = \begin{bmatrix} \nicefrac{B_{F,x}(t)}{G_x} \\ \nicefrac{B_{F,y}(t)}{G_y} \\ \nicefrac{B_{F,z}(t)}{G_z} \end{bmatrix} + \begin{bmatrix} 0 \\ 0 \\ \frac{B_p}{G_z}cos(2\pi f_d t) \end{bmatrix}
	\label{eq:3d_trajectory}.
\end{equation}
Defining $R_{s,i}(t) = \frac{d}{dt}B_{F,i}(t)$ as the SR for the FF along each direction, the assumption in Eq.~\eqref{eq:basic_condition} can be generalized as:
\begin{equation}
	B_p 2\pi f_d  \gg \underset{t}{\text{max}}\big\{ \big|R_{s,z}(t) \big|\big\}, \enspace \forall t \in [0, T_s].
	\label{eq:speed_condition}
\end{equation}
Here, $T_s$ is the total scan time. 
For example, for $B_p=15$~mT, $f_d=10$~kHz, and $R_{s,z}=20$~T/s, Eq.~\eqref{eq:speed_condition} yields a ratio of 47 between the left- and right-hand sides of the inequality. Therefore, considering that 20~T/s is the safety limit for time-varying magnetic fields, we can conclude that the FFP speed is dominated by the DF for all practical purposes \cite{international2004medical}. 

\subsection{Weighted Least Squares TAURUS}
The frequency domain division in Eq.~\eqref{eq:frequency_division_tau_corrected} makes TAURUS susceptible to division by a small number or zero under the presence of noise and interference.
Therefore, we propose casting  Eq.~\eqref{eq:frequency_division_tau_corrected} as a weighted least squares (WLS) problem by first expressing it as a vector relation:
\begin{equation}
	\mathbf{a}\tau = \mathbf{b},
\end{equation}
where $\mathbf{a}, \mathbf{b} \in \mathbb{C}^{N \times 1}$  such that:
\begin{subequations}
	\label{eq:wls-ab}
	\begin{align}
		& \mathbf{a} = i2\pi k \Delta f \big(S^*_{\text{pos,c}} (k \Delta f)- S_{\text{neg,c}} (k \Delta f) \big), \\
		& \mathbf{b} = S^*_{\text{pos,c}} (k \Delta f) + S_{\text{neg,c}} (k \Delta f).
	\end{align}
\end{subequations}
Here, $k = 0, 1, ..., N-1$ with $N = f_s/\Delta f$, where $f_s$ is the sampling frequency and 
$\Delta f$ is the resolution in frequency domain. WLS-TAURUS can then be written as: 
\begin{equation}
	\tau = Re\{(\mathbf{a}^\text{H}\mathbf{Wa})^{-1} \mathbf{a}^\text{H}\mathbf{W}\mathbf{b}\},
	\label{eq:ls-wls}
\end{equation}
where $\mathbf{W} \in \mathbb{R}^{N \times N}$ is a diagonal weighting matrix with  ${W_{k,k}}= |S_{\text{pos,c}}(k \Delta f)|^2$, and the superscript $\text{H}$ denotes Hermitian transpose. WLS-TAURUS avoids the problem of division by a small number or zero, and is expected to provide robustness against noise.

\section{Methods}
\subsection{Simulations}
The MPI simulations were carried out using a custom toolbox in MATLAB (Mathworks, Natick, MA). The parameters were chosen to mimic the experiments performed on our in-house MPI system:   $(-4.8,~2.4,~2.4)$~T/m SF gradients in x-, y- and z-directions, DF along the z-direction with $f_d = 10$~kHz and $B_p = 15$~mT, creating a pFOV of $W_p = 12.5$~mm. A homogeneous receive coil along the z-direction was utilized. The core MNP diameter was assumed to be 25~nm. 
To simulate the continuous-time nature of the physical world, the MNP responses were first generated at 100 MS/s, and then downsampled to 2 MS/s to generate the MPI signals. For direct feedthrough filtering, a zero-phase finite impulse response (FIR) high-pass filter (HPF) with a cut-off frequency of 1.5$f_d$ was utilized.  
The details of each simulation are explained in detail below.

\subsubsection{Signal Replication for Increased TAURUS Performance}
In Eq.~\eqref{eq:frequency_division_tau_corrected}, directly using the SR-corrected positive and negative signals, $s_{\text{pos,c}}(t)$ and $s_{\text{neg,c}}(t)$, results in a poor frequency resolution of $\Delta f = 2f_d$ (i.e., the inverse of a half DF period). 
To improve the frequency resolution, each period of the SR-corrected signal was first replicated $N_{\text{rep}}$ times, then concatenated with the positive signal on the right side for $s_{\text{pos,c}}(t)$ and with the negative signal on the left side for $s_{\text{neg,c}}(t)$ to yield the extended versions. 
This procedure increases the frequency resolution to $\Delta f = 2f_d/(N_{\text{rep}}+1)$. Example extended signals and the corresponding $\tau (f)$ for $N_{\text{rep}}$ values of 0, 4, and 9 at $R_{s,z}=20$~T/s are given in Supplementary Materials Fig.~S1. Note that the underlying mirror symmetry feature still applies for these extended signals. Next, the performances of TAURUS and WLS-TAURUS were evaluated as a function of $N_{\text{rep}} \in [0,\enspace 10]$ for the noise-free case and for relatively low signal-to-noise ratios (SNRs) of 2 and 10. The estimation error as a function of $N_{\text{rep}}$, shown in Supplementary Materials Fig.~S2, indicates that $N_{\text{rep}}<3$ results in poor performance especially at SNR = 2. The performances of TAURUS and WLS-TAURUS both converge to a constant level for $N_{\text{rep}}>5$, independent of the noise level. Therefore, for the rest of this work, $N_{\text{rep}}=6$ was utilized. 


\subsubsection{Slew Rate Robustness Analysis}
The proposed SR correction in Eq.~\eqref{eq:corrected} incorporates a time-shift correction and an amplitude correction. First, the performances of the no correction, only amplitude correction, only time-shift correction, and full SR correction cases were evaluated for $R_{s,z}~\in~[0,~20]$~T/s, for both TAURUS and WLS-TAURUS. Next, the SR robustness of the proposed method was evaluated for $R_{s,z} \in [0,~20]$~T/s and $R_{s,x} \in [0,~20]$~T/s, where the estimation performances of the no correction and SR correction cases were compared, both using WLS-TAURUS. For these simulations, a point source MNP distribution with $\tau = 3~\mu$s positioned at the origin was utilized. 

\subsubsection{Noise Robustness Analysis}
Monte Carlo simulations were performed to compare the noise robustness of TAURUS and WLS-TAURUS, for $SNR \in [1,~20]$ and $R_{s,z}~\in~[0,~20]$~T/s. In each case, the number of repetitions for the Monte Carlo simulations was set to $10^4$ for $SNR \le 5$, and $10^3$ for $SNR > 5$ to ensure convergence of the mean estimation error. In each repetition, white Gaussian noise with zero mean was added to the MPI signal. The standard deviation (std) of the noise was set as the maximum signal intensity (before direct feedthrough filtering) of the noise-free signal divided by the targeted SNR level. A point source MNP distribution  with $\tau = 3~\mu$s positioned at the origin was utilized. 

\begin{figure}[!h]
	\centering
	\includegraphics[width=0.8\linewidth]{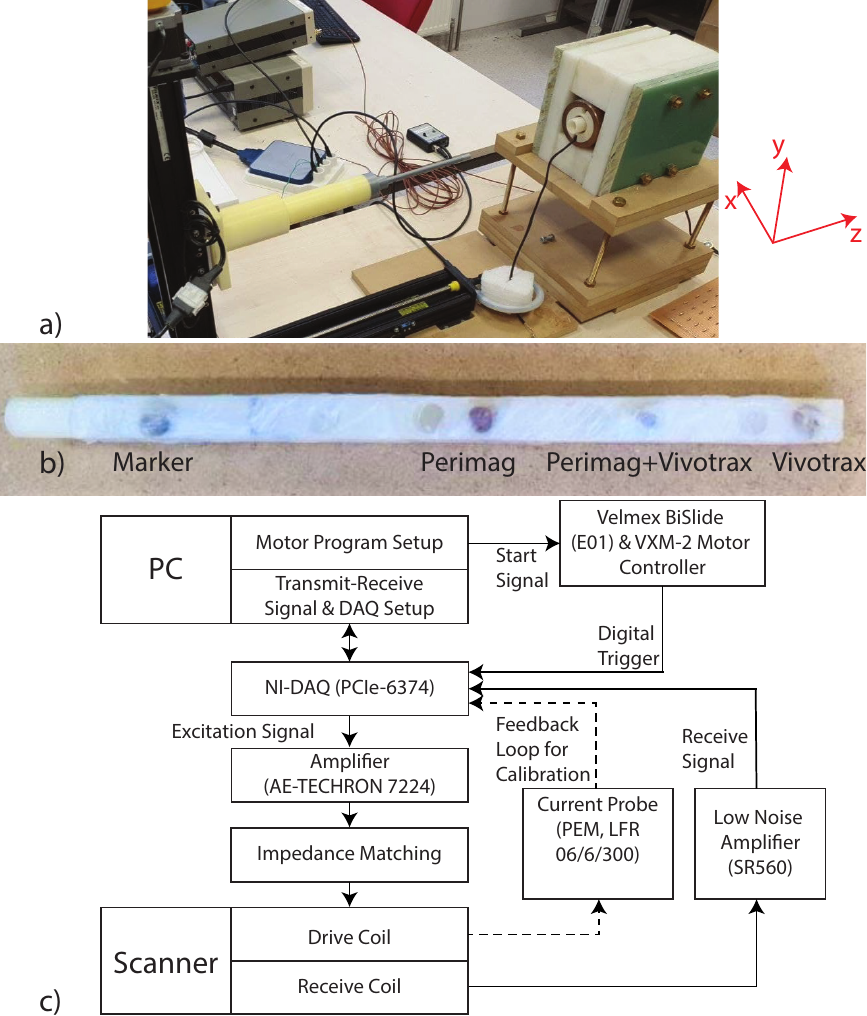}
	\caption{(a) In-house FFP MPI scanner with $(-4.8,~2.4,~2.4)$~T/m SF gradients. A three-axis linear actuator was utilized to move the FFP globally, instead of using FFs. (b) The imaging phantom contained Perimag, Vivotrax, their mixture, and a marker. (c) The flow diagram of the experimental process. A digital trigger from the VXM-2 motor controller was sent to the DAQ for synchronization.  
	}
	\label{fig:phantom_scanner_flow_diagram}
\end{figure}

\subsubsection{Color MPI Simulations}
\label{sec:full_mpi}
Finally, color MPI simulations were performed to evalute the performance of the proposed method. A phantom containing 6 MNP distributions with $\tau=$[2, 2.4, 2.8, 3.2, 3.6 4]~$\mu$s was created. Each MNP distribution had a size of 2$\times$2~mm$^2$. A 5$\times$6~cm$^2$ FOV in the x-z plane was scanned using 3 different trajectories: 

\textbf{Piecewise Trajectory (PWT):} The FFs along the x- and z-directions were stepped to cover the FOV at 100$\times$100 points in the x-z plane, providing 95.2\% overlap between the consecutive pFOVs along the z-direction. An MPI signal of 10~ms duration was simulated at each point, resulting in an active scan time of 100~s for the entire trajectory.

\textbf{Line-by-line Trajectory (LLT):} A linearly ramping FF in the z-direction was utilized with $R_{s,z} = 2$~T/s to cover each line in 72~ms, with 99.3\% overlap between the consecutive pFOVs along the z-direction. The FF along the x-direction was stepped to cover the x-direction at 100 equally spaced lines, resulting in an active scan time of 7.2~s. 

\textbf{2D Triangle Trajectory (2DTT):} To cover the 2D FOV continuously, a linearly ramping FF in the z-direction was utilized. A triangle wave FF was applied along the x-direction, formulated as follows:
\begin{equation}
	x(t) = \frac{\text{FOV}_x}{\pi}sin^{-1}\big(sin(2\pi f_\mathsf{T} t)\big), \enspace t\in [0, T_s).
	\label{eq:complex_rastered_ffp}
\end{equation}
Here, $f_\mathsf{T} = R_{s,x}/(2\text{FOV}_xG_x)$ is the frequency of the triangle wave and $T_s=\text{FOV}_z/(R_{s,z}/G_z)$. Here, $R_{s,z} = 0.01$~T/s and $R_{s,x} = 1$~T/s were chosen to densely cover the whole FOV. There was no explicit overlap among the neighboring pFOVs due to the continuous movement along the x-direction. The resulting active scan time was 14.4~s.

Note that the active scan times listed above do not include the idle times of PWT and LLT during which the FFP position is stepped. For example, traversing each line back at the same SR of $R_{s,z} = 2$~T/s would automatically double the total scan time of LLT to 14.4~s. In contrast, 14.4~s active scan time for 2DTT is directly equal to the total scan time, as the whole FOV is scanned continuously in a single shot.

\subsection{Imaging Experiments}
Color MPI imaging experiments were performed on our in-house FFP MPI scanner shown in Fig.~\ref{fig:phantom_scanner_flow_diagram}.(a), using 3 different trajectories. This scanner had $(-4.8,~2.4,~2.4)$~T/m SF gradients in the x-, y- and z-directions, and featured a free bore size of 1.9~cm and air cooling to prevent system heating. Both the drive and receive coils were oriented along the z-direction. To minimize the direct feedthrough, the receive coil was designed as a tunable 3-section gradiometric coil. Instead of utilizing FF coils to globally move the FFP, this scanner used a three-axis linear actuator (Velmex BiSlide), 
with a maximum speed of 38.1~mm/s in all axes.
The other details of this scanner can be found in  \cite{muslu2018calibration,utkur2019relaxation}. 

The flow diagram of the experimental process is presented in Fig.~\ref{fig:phantom_scanner_flow_diagram}.(c). For precise synchronization of the signal transmission/reception and the linear actuator movement, a digital trigger was sent from the VXM-2 motor controller of the actuator to a data acquisition card (DAQ) (NI PCIe-6374). The DAQ card sent the DF signal to the power amplifier (AE Techron 7224), which then sent it to the drive coil via an impedance matching circuitry tuned to $f_d = 10$~kHz. A current probe (PEM LFR 06/6/300) was used for calibrating $B_p$. The received signal was pre-amplified using a low-noise amplifier (SRS SR560) and then sampled at 2 MS/s using the DAQ.

\subsubsection{Trajectory Specifications}
\begin{figure}[!h]
	\centering
	\includegraphics[width=0.8\linewidth]{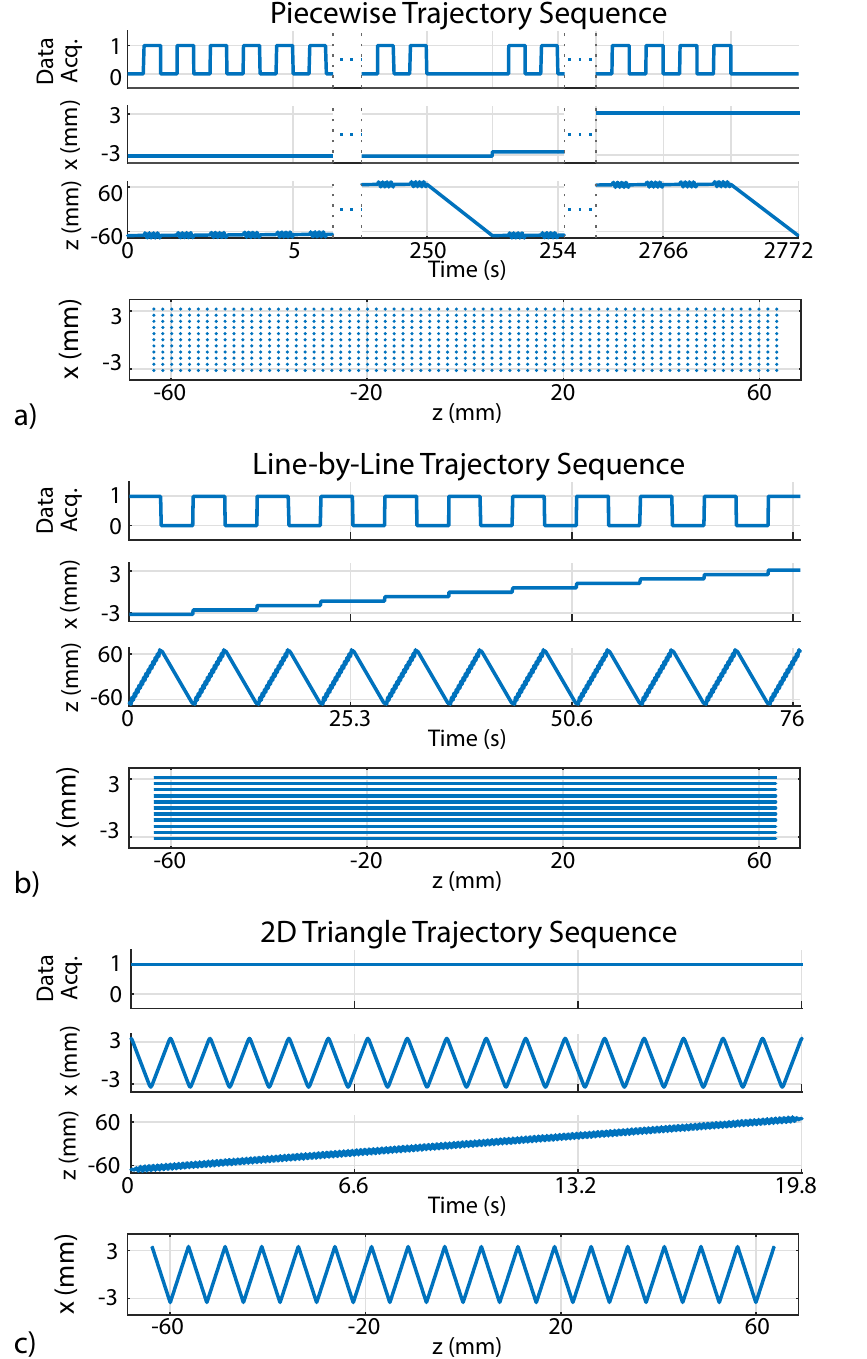}
	\caption{Trajectories used in the imaging experiments, with 0.7$\times$12.7~cm$^2$ FOV in the x-z plane. (a) PWT, with stepped x- and z- directions. 
		(b) LLT, with stepped x-direction and a linear motion along z-direction. 
		(c) 2DTT with a triangle-wave motion along x-direction and a linear motion along z-direction. 
	}
	\label{fig:trajectory_plot_column_2}    
\end{figure}

A 2D FOV of 0.7$\times$12.7~cm$^2$ in the x-z plane was scanned using 3 different trajectories, shown in Fig.~\ref{fig:trajectory_plot_column_2}. The experiments utilized $f_d = 10$~kHz and $B_p = 15$~mT, resulting in a pFOV size of $W_p=$12.5~mm. 

\textbf{PWT:} The FOV was divided into 11$\times$100 points in the x-z plane, providing 89.84\% overlap between the consecutive pFOVs. A 150~ms signal was acquired for each pFOV, resulting in an active scan time of 165~s. Due to the idle times needed for actuator motion, the total scan time was approximately 46~min.

\textbf{LLT:} The x-direction was divided into 11 equally spaced lines. For the z-direction, the actuator was moved continuously at its maximum speed of 38.1~mm/s, corresponding to $R_{s,z} = 0.091$~T/s for $G_z = 2.4$~T/m. This relatively low $R_{s,z}$ resulted in 99.97\% overlap between the consecutive pFOVs. Each line was covered in 3.46~s, resulting in an active scan time of 38~s. The total scan time was 76~s due to the idle times needed for the backward actuator movement along each line. 

\textbf{2DTT:} A continuous triangle wave movement with $R_{s,x} = 0.061$~T/s and $R_{s,z} = 0.03$~T/s was applied using Eq.~\eqref{eq:complex_rastered_ffp}. The active scan time was 19.8~s, with no idle time. 


For direct feedthrough compensation, baseline measurements were acquired before and after each line for PWT and LLT, and before and after the entire trajectory for 2DTT.


\subsubsection{Phantom Preparation}
As shown in Fig.~\ref{fig:phantom_scanner_flow_diagram}.(b), an imaging phantom was prepared with 3 different samples, containing Perimag (Micromod GmbH), Vivotrax (Magnetic Insight Inc.) and their equal volume mixture.  The samples  were placed with 2.3~cm center-to-center separations. Each sample had 3~mm diameter in the x-z plane and contained a total volume of 20~$\mu L$. Perimag (17~mg~Fe/mL undiluted concentration) was diluted 10.5 times to approximately equalize its signal level to that of Vivotrax (5.5~mg~Fe/mL undiluted concentration).
For computing signal timing (see Sec.~\ref{subsec:SignalTiming}), an additional marker sample containing 10~$\mu L$ undiluted Perimag was placed at a 4.6~cm separation from the leftmost sample. 


\subsubsection{Signal Preprocessing}
The received and baseline signals were individually low pass filtered using a zero-phase FIR filter with a cut-off frequency at 120~kHz to denoise the signal and remove the self-resonance of the receive coil at 280~kHz. 
Potential system drifts (e.g., due to heating or vibration) can cause slight delays between these two signals. This delay was negligibly small for PWT due to the short DF signals separated by extensive idle times that allowed system cooling. For LLT and 2DTT, the delay was more prominent due to the longer DF signals. The relative delay was computed by first upsampling the received and baseline signals at 100 MHz sampling rate, followed by a cross-correlation operation. The baseline signal was then time-shifted and subtracted from the received signal. Finally, the fundamental harmonic was filtered out using a zero-phase FIR HPF with a cut-off frequency of 1.5$f_d$.

\subsubsection{Digital Fine-Tuning of Signal Timing}
\label{subsec:SignalTiming}
TAURUS requires a precise (sub-sample level) adjustment of signal timing, as unaccounted delays can cause a bias in $\hat{\tau}$. The signal timing, $t_i$, of the preprocessed signal was fine tuned using the marker signal. To achieve sub-sample precision, first the signal was upsampled at 100~MHz sampling rate. The tuning process took advantage of the underlying mirror symmetry: when $r_{\tau}(t)$ computed at the correct $t_i$ is used to deconvolve the signal, the mirror symmetry should be achieved \cite{muslu2018calibration}, i.e.,
\begin{equation}
	\hat{t}_{i} = \underset{t_{\text{i}}}{\text{argmin}} \enspace \text{MSE} ( s_{\text{pos,d}}(t), -s_{\text{neg,d}}(-t) ).
	\label{eq:validation}
\end{equation}
Here, $\hat{t}_{i}$ is the estimated signal timing that minimizes the mean square error (MSE) in mirror symmetry. 
In addition, $s_{\text{pos,d}}(t)$ and $s_{\text{neg,d}}(t)$ denote the positive and negative signals after deconvolution with $r_{\tau}(t)$ computed at $t_i$, respectively.  
Finally, the estimated $\hat{t}_{i}$ was applied to the entire signal. 

\subsection{Color MPI Image Reconstruction}
\label{sec:Color MPI Image Reconstruction}
Color MPI image reconstruction was composed of three steps: MPI image reconstruction, $\hat{\tau}$ map reconstruction, and color overlay formation. 

\textbf{MPI Image Reconstruction:} The MPI images for PWT and LLT were reconstructed using Partial FOV Center Imaging (PCI) reconstruction \cite{kurt2020partial}. PCI requires the pFOV centers to be aligned on a line and to have a high overlap percentage, and hence is suitable for PWT and LLT. Since 2DTT does not have any explicit overlap between pFOVs, PCI is not directly applicable. Taking advantage of the continuous signal acquisition of this trajectory, Harmonic Dispersion X-space (HD-X) reconstruction was utilized \cite{kurt2020multi}. Since the data points in 2DTT were on a non-Cartesian grid, an automated gridding algorithm for non-Cartesian MPI reconstruction was utilized on the HD-X data \cite{ozaslan2019fully}. This gridding algorithm automatically tuned all reconstruction parameters, including the grid size. To reduce the blurring introduced by the gridding operation, the Cartesian output image was deconvolved with the gridding kernel using the Lucy-Richardson deconvolution. The final reconstructed images had 500$\times$600 pixels for the simulations and 200$\times$2133 pixels for the imaging experiments. 

\textbf{$\hat{\tau}$ Map Reconstruction}: After SR correction of each DF period, $\hat{\tau}$ was estimated using WLS-TAURUS with $N_{\text{rep}}=6$. 
The estimated $\hat{\tau}$ values were placed on the corresponding pFOV center locations to form a $\hat{\tau}$ map. For 2DTT, the  above-mentioned gridding algorithm 
was utilized to grid the $\hat{\tau}$ values to a Cartesian grid \cite{ozaslan2019fully}. In the background regions, $\hat{\tau}$ can have unexpectedly high or low values due to low SNR. To ensure that these outlier $\hat{\tau}$ values do not contaminate the $\hat{\tau}$ map,
the gridding kernel was reduced to one-fourth of that used during image reconstruction. Finally, the $\hat{\tau}$ map was interpolated to match the size of the MPI image. To suppress the noise-like $\hat{\tau}$ values in the background regions, the $\hat{\tau}$ map was multiplied with a binary mask of the MPI image, with a threshold of 10\% of the maximum pixel intensity.


\textbf{Color Overlay Formation}: The $\hat{\tau}$ maps were overlayed with the reconstructed MPI images by multiplying the RGB pixel intensities of the two images, channel by channel. 

\begin{figure}[!h]
	\centering
	\includegraphics[width=0.8\linewidth]{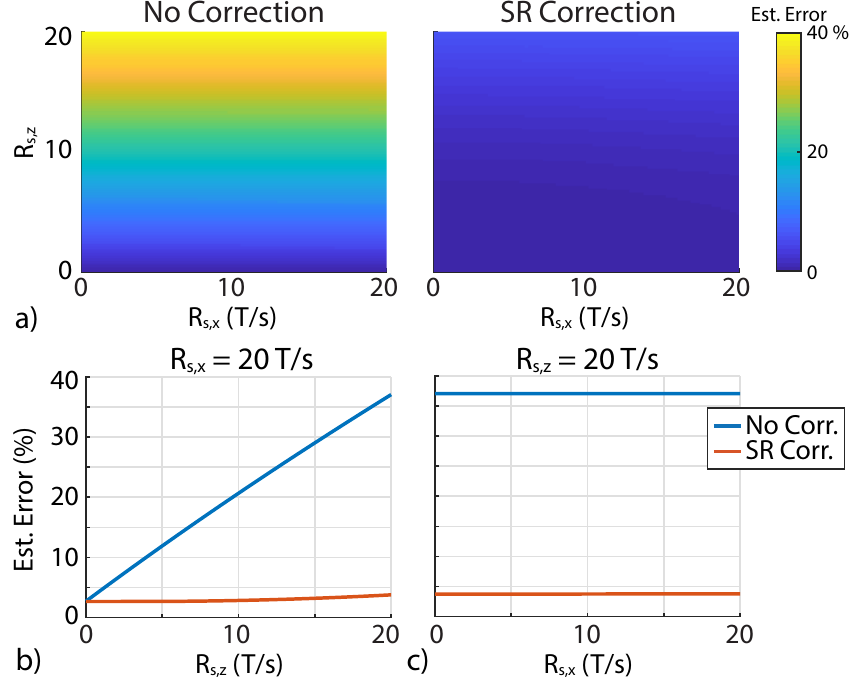}
	\caption{SR robustness results. The estimation error (a) as a function of $R_{s,x}$ and $R_{s,z}$ for no correction and SR correction, (b) as a function of $R_{s,z}$ for $R_{s,x} = 20$~T/s, and (c) as a function of $R_{s,x}$ for $R_{s,z} = 20$~T/s. SR correction reduces the estimation error to below 3.6\% at all $R_{s,z}$ and $R_{s,x}$ values.
	}
	\label{fig:combined_error_graph}
\end{figure}

\section{Results}
\subsection{SR Robustness Results}
The slew rate robustness of the proposed method is presented in Fig.~\ref{fig:combined_error_graph}.(a) as a function of $R_{s,z}$ and $R_{s,x}$, ranging between 0-20~T/s. Here, both the no correction and SR correction results utilized WLS-TAURUS. 
These results indicate that the estimation error does not depend on $R_{s,x}$, but is rather dominated by $R_{s,z}$ for the no correction case, reaching 37\% error at $R_{s,z}=20$~T/s. SR correction successfully reduces the estimation error to below 3.6\% at all $R_{s,z}$ and $R_{s,x}$ values tested. 
Figure~\ref{fig:combined_error_graph}.(b) shows the estimation error at $R_{s,x}=20$~T/s as a function of $R_{s,z}$. Even at this large $R_{s,x}$, SR correction maintains a low estimation error at all $R_{s,z}$ values tested. Next, Fig.~\ref{fig:combined_error_graph}.(c) shows the estimation error at $R_{s,z}=20$~T/s, displaying a less than 0.05\% increase as a function of $R_{s,x}$ for both the no correction and SR correction cases.

The component-wise performances of the proposed SR correction method were also evaluated for $R_{s,z}$ ranging between 0-20~T/s (see Supplementary Materials Fig.~S3). Amplitude correction alone has minimal effect on the estimation performance with respect to the no correction case. In contrast, time-shift correction alone provides a performance improvement that is only slightly lower than the full SR correction case.

\subsection{Noise Robustness Results}
\label{sec:Noise Robustness Analysis}
\begin{figure}[!h]
	\centering
	\includegraphics[width=0.8\linewidth]{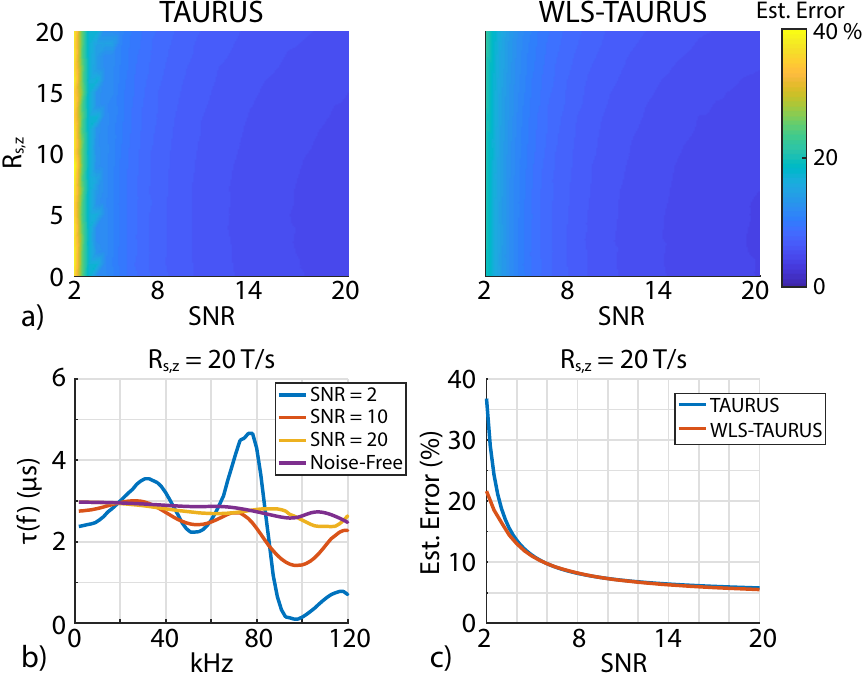}
	\caption{Noise robustness results. (a) The estimation error for TAURUS and WLS-TAURUS with respect to SNR and $R_{s,z}$. (b) $\tau (f)$ for 4 different SNR levels at $R_{s,z} = 20$~T/s. (c) TAURUS vs. WLS-TAURUS at $R_{s,z} = 20$~T/s. WLS-TAURUS shows improved robustness, especially for $\text{SNR} < 5$.}
	\label{fig:noise_performance_surf_plot}
\end{figure}

Noise robustness results for SR-corrected TAURUS and WLS-TAURUS are given in Fig.~\ref{fig:noise_performance_surf_plot} for SNR ranging between 2-20 and $R_{s,z}$ ranging between 0-20~T/s. This analysis did not include $R_{s,x}$, as Fig.~\ref{fig:combined_error_graph} indicated that $R_{s,x}$ does not have an impact on the estimation error. 
As seen in Fig.~\ref{fig:noise_performance_surf_plot}.(a), the estimation error after SR correction is almost independent of $R_{s,z}$, but depends largely on the noise level. At high SNR levels, WLS-TAURUS provides a relatively small improvement over TAURUS. At low SNR levels, TAURUS shows high estimation errors, particularly for $\text{SNR} < 5$. In contrast, WLS-TAURUS maintains robustness against noise even at $\text{SNR} = 2$.

\begin{figure}[!h]
	\centering
	\includegraphics[width=0.8\linewidth]{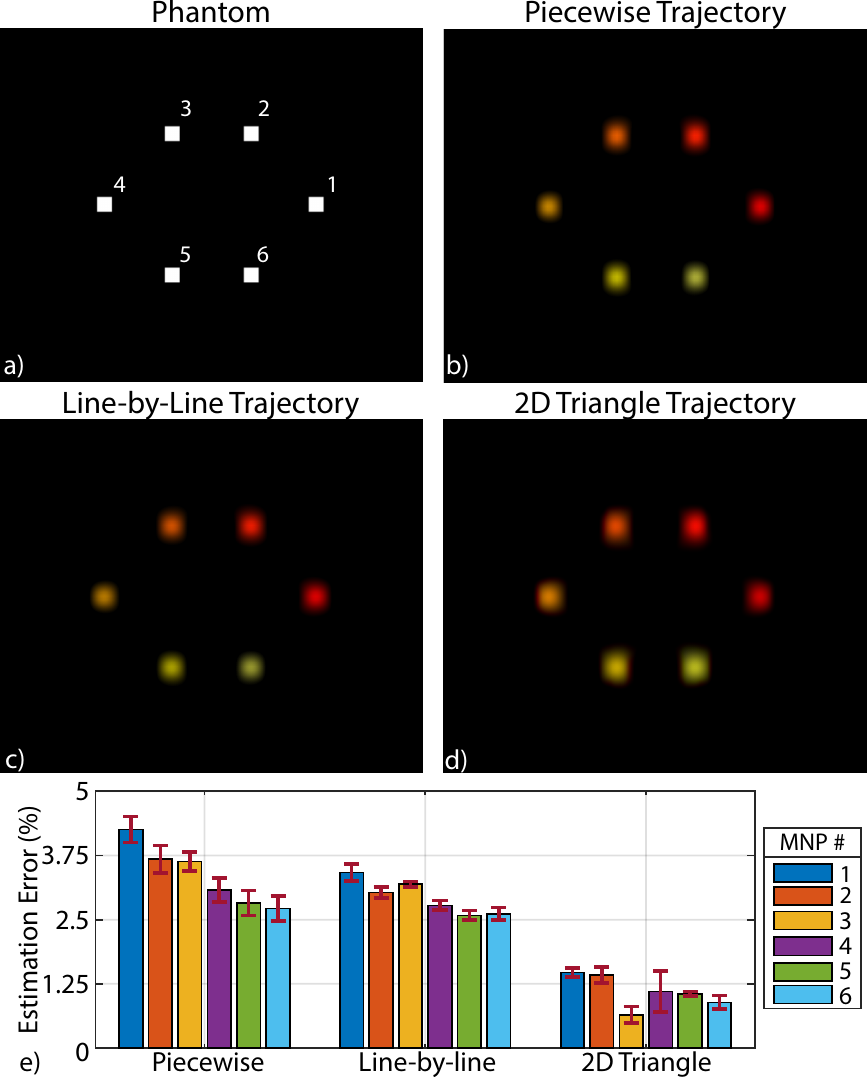}
	\caption{Color MPI simulation results for 3 different trajectories. (a) The digital phantom with $\tau$ between 2-4~$\mu$s. The color overlays for (b) PWT, (c) LLT, and (d) 2DTT, showing comparable performance despite their vastly different speeds. (e) The estimation performances, where the error bars denote the mean and std of the estimation error in $\hat{\tau}$. 2DTT shows improved estimation performance due to its dense coverage.}
	\label{fig:combined_without_colormap_final_column}
\end{figure}


In Fig.~\ref{fig:noise_performance_surf_plot}.(b), SR-corrected $\tau (f)$ in Eq.~\eqref{eq:frequency_division_tau_corrected} is plotted at 4 different SNR levels at $R_{s,z} = 20$~T/s. At $\text{SNR} = 2$, $\tau (f)$ shows large ripple-like deviations from the ideal value of $\tau = 3~\mu$s. These ripples decrease at high SNR and converge to the noise-free case. As explained before, these ripples increase with $R_{s,z}$. Hence, $\tau (f)$ shown for $\text{SNR} = 2$ and $R_{s,z} = 20$~T/s corresponds to the most challenging scenario.

Next, Fig.~\ref{fig:noise_performance_surf_plot}.(c) shows a direct comparison of TAURUS and WLS-TAURUS at $R_{s,z} = 20$~T/s as a function of SNR, where the improved noise robustness of WLS-TAURUS is clearly visible for $\text{SNR} < 5$. At $\text{SNR} = 2$, the estimation errors for TAURUS and WLS-TAURUS are 37\% and 21\%, respectively. 
At $\text{SNR} = 20$, the estimation errors fall down to 5.5\% and 5.0\% for TAURUS and WLS-TAURUS, respectively. In comparison, the estimation errors are 3.8\% and 3.6\% in the noise-free case for TAURUS and WLS-TAURUS, respectively (see Supplementary Materials Fig.~S3).




\subsection{Color MPI Simulation Results}

Figure~\ref{fig:combined_without_colormap_final_column} shows the color MPI simulation results for 3 different trajectories. The digital phantom used in these simulations is displayed in Fig.~\ref{fig:combined_without_colormap_final_column}.(a), with $\tau$ values between 2-4~$\mu$s. In Fig.~\ref{fig:combined_without_colormap_final_column}.(b)-(d), the color overlays show comparable performance for PWT, LLT, and 2DTT, despite their vastly different speeds. The estimation performances are compared quantitatively in Fig.~\ref{fig:combined_without_colormap_final_column}.(e), where the error bars denote the mean and std of the estimation errors in $\hat{\tau}$ within the 2$\times$2~mm$^2$ region for each MNP distribution. As seen in these results, 2DTT has the best performance, whereas LLT shows slightly improved performance over PWT. The mean estimation errors across all MNPs were 3.4$\pm$0.2\%, 3$\pm$0.1\% and 1.1$\pm$0.1\% for PWT, LLT, and 2DTT, respectively. These results directly reflect the effect of trajectory density and $\hat{\tau}$ map fidelity. While LLT is denser than PWT along the z-direction, both LLT and PWT are relatively sparse along the x-direction. In contrast, 2DTT provides a dense coverage along both the x- and z-directions, resulting in improved estimation performance.

\begin{figure}[!h]
	\centering
	\includegraphics[width=0.8\linewidth]{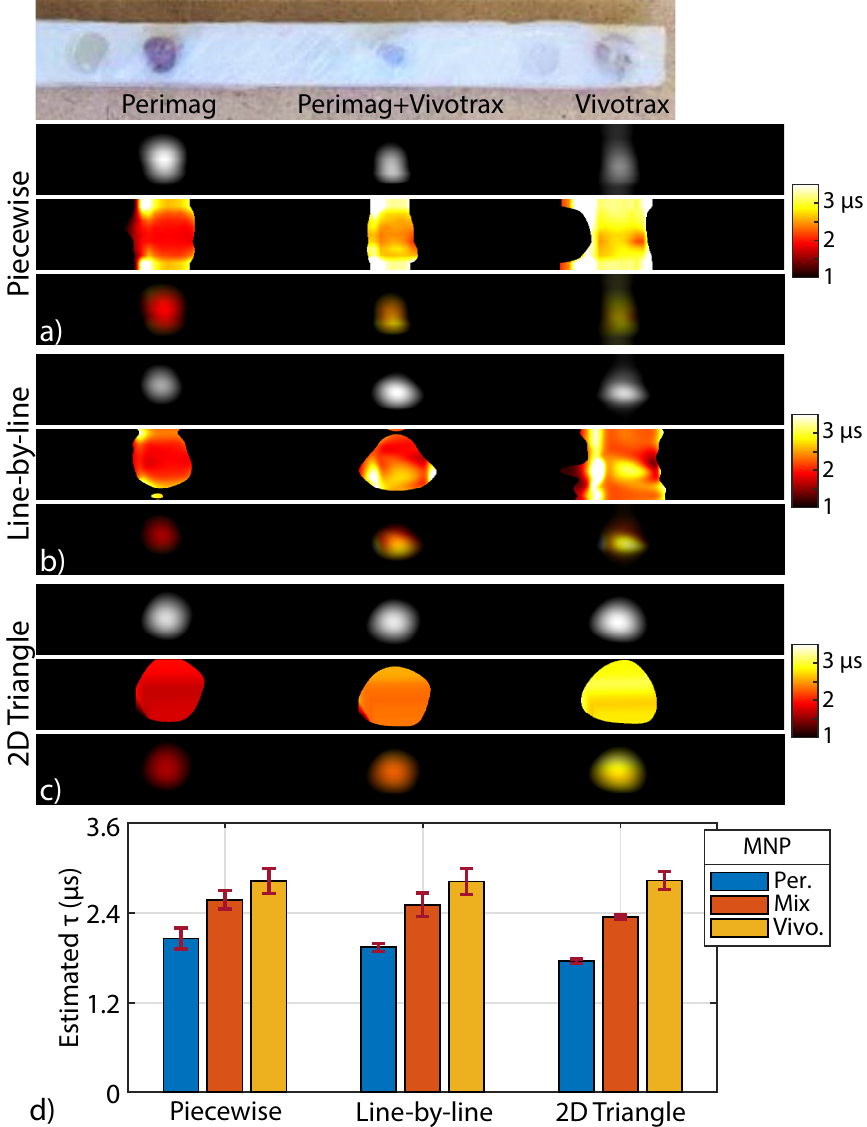}
	\caption{Imaging experiment results. The reconstructed MPI images, $\hat{\tau}$ maps, and color overlays for (a) PWT, (b) LLT, and (c) 2DTT. $\hat{\tau}$ map for 2DTT is visibly smoother. (d) The estimated $\hat{\tau}$ for each sample, where the  error  bars denote  the  mean  and  std within the FWHM region of the respective MPI image for that sample and trajectory. The displayed FOV is 0.7$\times$7.3~cm$^2$.}
	\label{fig:combined_colormap_wo_marker_deconv}
\end{figure}

\subsection{Imaging Experiment Results}
Figure~\ref{fig:combined_colormap_wo_marker_deconv}.(a)-(c) displays the imaging experiment results, showing the reconstructed MPI images, $\hat{\tau}$ maps, and color overlay images for 3 different trajectories. 
For PWT and LLT, $\hat{\tau}$ shows a large variation along the x-direction due to the low trajectory density in that direction. For 2DTT, $\hat{\tau}$ map is visibly smoother with reduced variations along both x- and z-directions, as this trajectory is considerably denser than the other two trajectories. Since Vivotrax has worse full-width at half-maximum (FWHM) resolution than Perimag, its $\hat{\tau}$ values extend in a wider region along the z-direction. 

The estimated $\hat{\tau}$ for each sample are given as bar plots in Fig.~\ref{fig:combined_colormap_wo_marker_deconv}.(d). Here, the  error  bars denote  the  mean  and  std  of  $\hat{\tau}$ of each sample within the FWHM region of the respective MPI image for that sample and trajectory. For PWT, the $\hat{\tau}$ values were 2.06$\pm$0.14~$\mu$s, 2.58$\pm$0.13~$\mu$s, and 2.83$\pm$0.16~$\mu$s for Perimag, mixture, and Vivotrax, respectively. For LLT, the $\hat{\tau}$ values were 1.94$\pm$0.05~$\mu$s, 2.51$\pm$0.16~$\mu$s, and 2.82$\pm$0.17~$\mu$s for Perimag, mixture, and Vivotrax, respectively. For 2DTT, the $\hat{\tau}$ values were 1.78$\pm$0.03~$\mu$s, 2.36$\pm$0.02~$\mu$s, and 2.83$\pm$0.12~$\mu$s for Perimag, mixture, and Vivotrax, respectively. These results indicate that $\hat{\tau}$ for Vivotrax is consistent, showing less than 1\% variation across trajectories. This consistency is potentially due to the worse FWHM resolution of Vivotrax, which provides a paradoxical robustness against the low densities of PWT and LLT along the x-direction. In addition, $\hat{\tau}$ for Perimag is also consistent between PWT and LLT, with only 5.8\% difference, as their densities match along the x-direction. In contrast, $\hat{\tau}$ for Perimag for 2DTT is approximately 13.6\% and 8.2\% lower than for PWT and LLT, respectively. These differences potentially stem from the substantially increased density of 2DTT along the x-direction. 
For each trajectory, $\hat{\tau}$ for the mixture closely matches the average of $\hat{\tau}$ for Perimag and Vivotrax, as expected \cite{muslu2018calibration}.

Overall, these results indicate that 2DTT provides reliable $\hat{\tau}$ estimations thanks to its dense coverage, while also providing the shortest scan time among the tested trajectories.


\section{Discussion}
In this work, we have proposed and experimentally demonstrated techniques that enable relaxation-based color MPI for rapid and multi-dimensional trajectories. The proposed SR correction method provides excellent robustness against FF-induced distortions for a wide range of SR values. Importantly, this method solely depends on the system and trajectory parameters. In addition, using a weighted least squares approach, WLS-TAURUS, successfully improves the noise robustness.


\subsection{Alternative SR Correction Approaches}
The proposed algorithm applies $\Delta t$ time-shift and $\alpha_{\Delta t}$ scaling on the negative signal to align it with the positive signal. An alternative approach could be to apply $\Delta t /2$ time-shift and  $\sqrt{\alpha_{\Delta t}}$ scaling on the negative signal and $- \Delta t /2$ time-shift and  $1/\sqrt{\alpha_{\Delta t}}$ scaling on the positive signal, so that they meet at the center of the half period. This approach would provide identical performance to the proposed SR correction, given that SR-induced shifts are within 1\% of the DF period, even at $R_{s,z} = 20$~T/s.
Another approach could be to fully correct for the time-varying nature of the FF-induced distortions, so that the corrected $s_{\text{adiab}} (t)$ attains perfect mirror symmetry. However, such a correction would essentially stretch or compress one (or both) of the half-cycle signals, perturbing the temporal properties of the relaxation effect in Eq.~\eqref{eq:time_domain_eqs}. In contrast, the proposed SR correction preserves the relaxation effects within individual half cycles. 

\subsection{SR Robustness}
\label{sec:Amount of Distortion of Symmetry}
The simulations presented in this work were based on our in-house MPI scanner, with both the drive and receive coils along the z-direction. As shown in Fig.~\ref{fig:combined_error_graph}, the SR correction is very robust against $R_{s,z}$ and does not depend on $R_{s,x}$. The latter effect partially stems from the fact that the collinear PSF is relatively wide along the x-direction \cite{goodwill2011multidimensional}. 
For the simulation parameters in this work, the collinear PSF has 3.7~mm and 7.5~mm FWHM along the z- and x-directions, respectively. Even for $R_{s,x} = 20$~T/s, the FFP moves approximately 0.2~mm along the x-direction in half the DF period. 
The fact that this movement is considerably smaller than the FWHM along the x-direction is the fundamental reason behind the independence from $R_{s,x}$ (see Supplementary Materials Fig.~S4 for a visual explanation). For the same reasons, we expect $R_{s,y}$ to have a similarly negligible effect.

\subsection{Effects of MNP Diameter}
According to the Langevin theory,  increasing the MNP diameter or the SF gradients decreases the FWHM of the PSF in all directions \cite{goodwill2011multidimensional}. Regardless, the MNP diameter should not have any major effects on the $\tau$ estimation performance. First of all, SR correction is independent of the MNP parameters. Secondly, in terms of resolution, increasing the MNP diameter is equivalent to increasing $G_z$ instead. The trajectory can still be kept identical if $B_p$ and $R_{s,z}$ are increased at the same rate as $G_z$. Because the assumption in Eq.~\eqref{eq:speed_condition} remains unchanged, the estimation performance with respect to $R_{s,z}$ should not be affected. In addition, unless the FWHM along the x-direction becomes very small, 
the estimation performance with respect to $R_{s,x}$ should not be affected, either. As mentioned above, $R_{s,x} = 20$~T/s moves the FFP by $\sim$0.2~mm in half the DF period. Even in theory, reaching a comparable FWHM along the x-direction would require an effective MNP diameter greater than 80~nm. 




\subsection{Resolution of $\tau$ Map}
\label{sec:Resolution of the Relaxation Map}
When creating the $\hat{\tau}$ map, a single $\hat{\tau}$ is estimated for each period of the signal and assigned to the center of the corresponding pFOV. Consequently, the resolution of $\hat{\tau}$ map is directly proportional to the trajectory density. For PWT, improving the resolution requires increasing the number of stepped points along both the x- and z-directions, increasing the scan time quadratically for a 2D FOV. For LLT, the resolution along the continuously scanned z-axis is sufficiently high and can be further improved by reducing $R_{s,z}$ (e.g., even at $R_{s,z} = 20$~T/s, the resolution along the z-axis is 0.83~mm for the parameters in this work). However, improving the resolution along the x-direction requires more lines to be scanned, which would increase the scan time linearly for a 2D FOV. In 2DTT, the resolution is considerably improved along the x-direction, but is position dependent. For a fixed scan time and FOV, $R_{s,z}$ and the number of pFOVs are also fixed. Then, a trade-off between the resolutions in the x- and z-directions can be achieved by adjusting $R_{s,x}$. By increasing the scan time and reducing $R_{s,z}$, the overall resolution of $\hat{\tau}$ map can be further improved in both directions. 



It should be noted that controlling the trajectory density in a systematic fashion is challenging for 2DTT, as the trajectory parameters are coupled. For example, doubling $f_d$ alone will double the density on the exact same trajectory. The same effect can also be achieved if $R_{s,x}$ and $R_{s,z}$ are both reduced to half with a doubling in scan time to cover the same FOV. However, in both cases, the off-trajectory points will not benefit from this seeming increase in density. In contrast, if we reduce only $R_{s,z}$ to half with a doubling in scan time, the trajectory itself will have twice as many triangles covering the same FOV. This time, the increased density will benefit the entire scanned FOV. Furthermore, the differences in scan time also need to be considered for a fair comparison of trajectories with respect to the noise level. Considering the coupling between the scan parameters, the effects of trajectory density on $\hat{\tau}$ map fidelity must be analyzed thoroughly with an appropriate definition of density, which remains an important future work.

\subsection{Extension to Other Rapid MPI Trajectories}
\label{sec:Rapid MPI Trajectories}
This work utilized constant SRs in both the simulations and the experiments. In fact, a constant but high SR (e.g., 20~T/s) is the worst-case scenario for FF-induced distortions. If the assumption in Eq.~\eqref{eq:speed_condition} is satisfied, extending the proposed method to time-varying SRs should not pose a challenge. 


In this work, a 1D DF was utilized. Trajectories with multi-dimensional DFs, such as the Lissajous trajectory, can cover a 2D/3D pFOV in a relatively short scan time \cite{knopp2008trajectory}. For such trajectories, the assumption in Eq.~\eqref{eq:speed_condition} has to be satisfied for each DF axis. Note that TAURUS requires a back-and-forth scanning. Therefore, the multi-dimensional DFs may need to be applied twice, once forward and once backward \cite{muslu2018calibration}. Extending TAURUS to such multi-dimensional DFs remains an important future work.

\section{Conclusion}
In this work, we have proposed a novel SR correction method to compensate for the FF-induced distortions in the underlying mirror symmetry of the MPI signal. The proposed method depends only on the system and scanning parameters, and enables high-fidelity relaxation map estimations via TAURUS for rapid and multi-dimensional trajectories. The performance is further boosted via a weighted least squares approach. The results show robustness against a wide range of SRs and noise, together with orders of magnitude reduction in scan time. The proposed rapid relaxation mapping method will have important applications in developing the functional imaging capabilities of MPI. 


\section{Acknowledgments}

A preliminary version of this work was presented at the 10$^{th}$ International Workshop on Magnetic Particle Imaging (IWMPI 2020). The authors would like to thank Dr. Mustafa Utkur for valuable discussions and feedback.

\bibliographystyle{IEEEtran}
\bibliography{references}

\end{document}